# First-principles theory of ionic thermoelectricity

Byeoksong Lee and Joonggoo Kang[*]

Department of Physics and Chemistry, DGIST, Daegu 42988, Republic of Korea

Symmetry plays a crucial role in shaping the theories of fundamental forces. For example, general covariance—the equivalence of all possible coordinate systems of spacetime—dictates the law of gravity. Here, we extend this concept to nonequilibrium thermodynamics by developing a theory of ionic thermoelectricity ("thermoelectricity without electrons") in electronically gapped ionic conductors. Within the Green-Kubo formalism, we show that energy gauge invariance—the equivalence of all possible ways of distributing energy among atoms—primarily determines the expressions for ionic thermoelectric coefficients. This symmetry-dictated theory is generally applicable, regardless of specific ion transport mechanisms, and provides a rigorous conceptual and computational framework for describing ionic thermoelectricity from first principles.

Thermoelectricity (TE), present in all electrically conductive materials, can be broadly classified into electronic and ionic TE based on the type of charge carriers (Fig. 1). Ionic TE arises from the motion of ions (thermodiffusion effect[1–4]) in electronically gapped systems such as electrolytes[5–7], ion gels[8,9], and free nerve endings[2,10]. There is growing interest in harnessing TE processes in ionic energy materials. For example, a figure of merit as high as $ZT = 6.1$ has been reported for cation-doped ion gels[9]. Giant thermopower on the order of 10 mV/K has been observed in cellulose ionic conductors[6] and polymer gels[8]. Recently, the Peltier coefficients of lithium-ion electrolytes were precisely measured[7]. However, the fundamental understanding of ionic TE still lags behind its electronic counterpart, hindering the development of efficient ionic TE materials.

The underlying principles differ between electronic and ionic TE, involving the Fermi-Dirac (FD) statistics for the former and the Maxwell-Boltzmann (MB) statistics for the latter. According to the Mott relation[11], electronic Seebeck coefficients $S_e$ at temperature $T$ can be expressed as

$$S_e = \frac{1}{T}\frac{\langle \varepsilon - \mu \rangle_{FD}}{\langle q \rangle_{FD}}, \quad (1)$$

where $\langle \varepsilon - \mu \rangle_{FD} = \frac{\int (\varepsilon-\mu)\sigma(\varepsilon)\left(-\frac{df_{FD}}{d\varepsilon}\right)d\varepsilon}{\int \sigma(\varepsilon)\left(-\frac{df_{FD}}{d\varepsilon}\right)d\varepsilon}$ is the average of the electron energy $\varepsilon$, referenced to the chemical potential $\mu$, based on the energy-dependent conductivity $\sigma(\varepsilon)$ and the FD distribution $f_{FD}(\varepsilon) = \frac{1}{e^{(\varepsilon-\mu)/k_B T}+1}$, while $\langle q \rangle_{FD} = -|e|$ for electron charge $e$. However, for ionic Seebeck coefficients, it remains unknown what would be an MB analogue of equation (1).

Simple transport models[3,4] provide useful insights into estimating ionic thermopower in redox-free ionic systems. However, first-principles calculations of TE properties in real materials require a microscopic theory that is generally applicable, regardless of specific ion transport mechanisms. The Green-Kubo (GK) theory of linear response[12–15] is a computational framework for extracting transport coefficients from energy, charge, and particle number fluxes in *equilibrium* molecular dynamics (MD) simulations. For an $N$-atom system of volume $\Omega$, the electric charge flux $\boldsymbol{J}^C(t)$ at time $t$ is given by[16]

$$\boldsymbol{J}^C(t) = \frac{1}{\Omega}\sum_{I=1}^{N} \boldsymbol{Z}_I^*(t)\cdot \boldsymbol{v}_I(t), \quad (2)$$

where $\boldsymbol{Z}_I^*$ and $\boldsymbol{v}_I$ are the atomic charge tensor and the velocity of atom $I$, respectively. From an MD trajectory $\boldsymbol{R}_I(t)$, the tensor components of $\boldsymbol{Z}_I^*$ (also known as the Born effective charge[17,18]) are given as $Z_{I\alpha\beta}^* = \Omega \frac{\partial P_\alpha}{\partial R_{I\beta}}$ for macroscopic polarization $\boldsymbol{P}$[19–21]. Similarly, the energy flux $\boldsymbol{J}^E(t)$ can be expressed in terms of local "atomic" energies carried by individual atoms:[22,23]

$$\boldsymbol{J}^E(t) = \frac{1}{\Omega}\sum_{I=1}^N \boldsymbol{E}_I^*(t) \cdot \boldsymbol{v}_I(t). \qquad (3)$$

However, unlike the Born effective charge $\boldsymbol{Z}_I^*$, the atomic energy tensor $\boldsymbol{E}_I^*$ in equation (3) is inherently ill-defined because decomposing total energy into atomic energies is an *ad hoc* process for interacting particle systems[22–27]. For example, when atoms interact through a pair potential, the distribution of the potential energy between the two atoms is arbitrary[25].

Here, we demonstrate how the problem with ill-defined atomic energies can be made into an *opportunity* for formulating a physical theory of ionic TE. Any candidate theory should yield well-defined, invariant transport coefficients, irrespective of how atomic energies are assigned. This requirement, hereafter referred to as *energy gauge invariance*, imposes significant constraints on the possible forms of a candidate theory. Indeed, we show that energy gauge invariance, combined with two other symmetries—Onsager's reciprocal relation[28,29] and the topological invariance of adiabatic charge transport[16]—inevitably leads to the expressions for ionic TE coefficients. The ambiguity arising from ill-defined atomic energies is resolved by renormalizing partial enthalpies in terms of the atomic energies themselves. As a proof of concept, we present numerical results for Cu$_2$S (Cu-ion liquid-like TE material[30,31]) and Li$_3$N (Li-ion solid-state electrolyte[32]).

**Theory**

The GK theory of linear response[12–15] describes a system's response to external perturbations in terms of the fluctuations of the unperturbed system. In an *n*-species system, the energy flux $J_x^E$ in direction $x$ and the species-specific particle number fluxes $\boldsymbol{J}_x^N = \left(J_{x,1}^N, J_{x,2}^N, \ldots, J_{x,n}^N\right)^{\mathrm{T}}$ are related to perturbations as $\begin{bmatrix} J_x^E \\ \boldsymbol{J}_x^N \end{bmatrix} = \begin{pmatrix} \Lambda^{EE} & \boldsymbol{\Lambda}^{EN} \\ \boldsymbol{\Lambda}^{NE} & \boldsymbol{\Lambda}^{NN} \end{pmatrix} \begin{bmatrix} \frac{\partial}{\partial x}\left(\frac{1}{k_B T}\right) \\ \frac{\partial}{\partial x}\left(-\frac{\boldsymbol{\mu}}{k_B T}\right) \end{bmatrix}$, where $T$ is temperature and $\boldsymbol{\mu} = (\mu_1, \ldots, \mu_n)^{\mathrm{T}}$ denotes the chemical potentials of *n* species. The scalar element $\Lambda^{EE}$ in the matrix is obtained from the energy flux $\mathcal{J}_x^E(t)$ in equation (3) as $\Lambda^{EE} = \Omega \int_0^\infty \langle \mathcal{J}_x^E(0) \mathcal{J}_x^E(\tau) \rangle d\tau$, where $\langle \cdots \rangle$ indicates an ensemble average in equilibrium MD for a time lag $\tau$. For the number fluxes $\mathcal{J}_{x,i}^N(t) = \frac{1}{\Omega} \sum_{I \in i} v_{Ix}(t)$ of species *i*, the $n \times 1$ vector $\boldsymbol{\Lambda}^{NE}$ has elements $\Lambda_i^{NE} = \Omega \int_0^\infty \langle \mathcal{J}_{x,i}^N(0) \mathcal{J}_x^E(\tau) \rangle d\tau$, and $\boldsymbol{\Lambda}^{EN} = (\boldsymbol{\Lambda}^{NE})^{\mathrm{T}}$. For species *i* and *j*, the element $\Lambda_{ij}^{NN}$ of the $n \times n$ block matrix $\boldsymbol{\Lambda}^{NN}$ is given by $\Lambda_{ij}^{NN} = \Omega \int_0^\infty \langle \mathcal{J}_{x,i}^N(0) \mathcal{J}_{x,j}^N(\tau) \rangle d\tau$, and $\Lambda_{ij}^{NN} = \Lambda_{ji}^{NN}$ due to time-reversal symmetry.

The energy flux $\boldsymbol{\mathcal{J}}^E(t)$ in equilibrium MD, a key component of the GK formalism, depends on the specific scheme used to assign atomic energies. From an equilibrium MD trajectory $\boldsymbol{R}_I(t)$ and associated kinetic $T_I(t)$ and potential $U_I(t)$ energies of atom *I*, the tensor components $E_{I\alpha\beta}^*(t)$ in equation (3) are given by

$$E_{I\alpha\beta}^* = (T_I + U_I)\delta_{\alpha\beta} + \sigma_{I\alpha\beta}, \qquad (4)$$

where $\delta_{\alpha\beta}$ is the Kronecker delta for Cartesian directions $\alpha$ and $\beta$, and $\sigma_{I\alpha\beta} = -\sum_{J=1}^N (R_{I\alpha} - R_{J\alpha}) \frac{\partial U_J}{\partial R_{I\beta}}$ is a virial term[22,23]. Throughout this paper, $I = 1, 2, \cdots, N$ denotes the atomic index in an *N*-atom system, while $i = 1, 2, \cdots, n$ represents the species index in an *n*-species system. While the total sums of $U_I(t)$ and $\sigma_{I\alpha\beta}(t)$ over atoms *I* are well-defined, the individual values of $U_I(t)$ and $\sigma_{I\alpha\beta}(t)$ are inherently ill-defined in interacting particle systems.

To illustrate this, we generated an ensemble of 100 machine-learned potentials (MLPs) for the system $\alpha$-Cu$_2$S (Fig. 2a and Methods). In an MLP scheme[33–38], which was originally developed as a force field for large-scale MD simulations[33], local atomic energies of atoms *I*

$$U_I = U_{\mathrm{MLP}}(x_I) \qquad (5)$$

are determined based on local atomic environments $x_I$ to satisfy the sum rule $U = \sum_{I=1}^N U_{\mathrm{MLP}}(x_I)$ for a total potential energy *U*. To systematically generate 100 MLPs of the same quality as force fields, we chose a mixed kernel model in Gaussian Approximation Potential[36,37]. All MLPs in the ensemble were prepared using the same kernel and training data. The only difference among the MLPs lies in the sparse Gaussian process regression (GPR) fitting[37]. While all MLPs possess the same quality as force fields, they yield different values for $U_I(t)$ and $\sigma_{I\alpha\beta}(t)$ from a given MD trajectory $\boldsymbol{R}_I(t)$. Figure 2b shows the time averages,

$$\bar{U}_i = \frac{1}{N_i}\sum_{I\in i}\langle U_I(t)\rangle \text{ and } \bar{\sigma}_i = \frac{1}{N_i}\sum_{I\in i}\langle \sigma_{I\alpha\alpha}(t)\rangle, \quad (6)$$

for $i =$ Cu and S in the cubic phase $\alpha$-Cu$_2$S, both following Gaussian distributions. (For anisotropic systems, $\bar{\sigma}_i$ should be denoted as $\bar{\sigma}_{i\alpha}$ as it depends on the direction $\alpha$.) Still, as illustrated by the straight lines in Fig. 2b, $\bar{U}_i$ and $\bar{\sigma}_i$ satisfy the sum rules for the total potential and virial energies, respectively: $\sum_{i=1}^n N_i \bar{U}_i = \langle U(t)\rangle$ and according to the virial theorem[39], $\sum_{i=1}^n N_i \bar{\sigma}_i = P\Omega - Nk_BT$ for volume $\Omega$ and pressure $P$.

Choosing a scheme for assigning atomic energies $U_I$ is akin to selecting a coordinate system in analytic geometry. All "coordinate" systems are equivalent if they preserve atomic forces $\boldsymbol{F}_I = -\nabla_I \sum_{J=1}^N U_J$. For example, we may consider a simple symmetry transformation (ref. [23]) involving species-specific, time-independent energy shifts (or "gauges") $\Delta_i$,

$$U_I(t) \mapsto U_I(t) + \Delta_{s(I)}, \quad (7)$$

where $s(I)$ denotes the species of atom $I$. This gauge transformation is "global" in the sense that $\Delta_{s(I)}$ is independent of the local atomic environments $x_I$, thereby preserving the virial terms $\sigma_{I\alpha\beta}$. In contrast, the switching between MLPs in Fig. 2 represents a "local" (and thus more general) transformation that depends on $x_I$, leading to different values of $\sigma_{I\alpha\beta}$.

The MLPs in the ensemble lead to different energy fluxes $\boldsymbol{J}^E(t)$ due to the arbitrariness of $U_I$ and $\sigma_{I\alpha\beta}$ in equation (4). To clarify its implications, we rewrite the linear response equation as

$$\begin{bmatrix}\tilde{J}_x^E\\ J_x^N\end{bmatrix} = \begin{pmatrix}\tilde{\Lambda}^{EE} & \tilde{\Lambda}^{EN}\\ \tilde{\Lambda}^{NE} & \Lambda^{NN}\end{pmatrix}\begin{bmatrix}-\frac{1}{k_BT^2}\frac{\partial T}{\partial x}\\ -\frac{1}{k_BT}\left(\frac{\partial \mu}{\partial x}\right)_T + \frac{\tilde{h}}{k_BT^2}\frac{\partial T}{\partial x}\end{bmatrix}, \quad (8)$$

where MLP-dependent quantities are marked with a tilde. The vector of external perturbations is obtained using the thermodynamic relation $\frac{\partial}{\partial x}\left(-\frac{\mu}{k_BT}\right) = -\frac{1}{k_BT}\left(\frac{\partial \mu}{\partial x}\right)_T + \frac{h}{k_BT^2}\frac{\partial T}{\partial x}$, where $\left(\frac{\partial \mu}{\partial x}\right)_T$ denotes the field-induced gradient of $\boldsymbol{\mu}$ at constant $T$. For enthalpy $H$, the vector $\boldsymbol{h} = (h_1, \ldots, h_n)^T$ represents the partial enthalpies $h_i = \left(\frac{\partial H}{\partial N_i}\right)_{T,P}$, satisfying the sum rule[40] $H = \sum_{i=1}^n N_i h_i$.

Two key points should be noted here. First, the number flux $J_x^N$ must remain invariant under different choices of MLPs. This requires that the partial enthalpies in equation (8) must be properly "renormalized" in terms of the atomic energies $U_I$ (i.e., $h_i \longrightarrow \tilde{h}_i$) so that the ill-defined parts of $\widetilde{\mathbf{\Lambda}}^{NE}$ and $\widetilde{\mathbf{h}}$ exactly cancel out, leaving $J_x^N$ invariant. Second, while $\tilde{J}_x^E$ is ill-defined, this is not problematic as what is measured is the heat flux $J_x^H$ rather than the energy flux $\tilde{J}_x^E$ itself. Therefore, the main challenge lies in constructing a well-defined heat flux $J_x^H$ from the ill-defined ingredients in equation (8) *through* the renormalization of $\widetilde{\mathbf{h}}$.

In the following, we first identify the heat flux $J_x^H$ in a form that remains invariant under the global gauge transformation in equation (7). Next, we discuss how partial enthalpies must be renormalized to ensure that $J_x^H$ remains invariant under local gauge transformations.

**Global gauge invariance.** Under the gauge transformation in equation (7), the atomic energy tensor in equation (4) transforms as $E^*_{I\alpha\beta}(t) \mapsto E^*_{I\alpha\beta}(t) + \Delta_{s(I)}\delta_{\alpha\beta}$. Hence, $\widetilde{\Lambda}^{EE} = \Lambda^{EE}(\mathbf{\Delta})$ and $\widetilde{\mathbf{\Lambda}}^{NE} = \mathbf{\Lambda}^{NE}(\mathbf{\Delta})$ in $\mathbf{\Lambda} = \begin{pmatrix} \widetilde{\Lambda}^{EE} & \widetilde{\mathbf{\Lambda}}^{EN} \\ \widetilde{\mathbf{\Lambda}}^{NE} & \mathbf{\Lambda}^{NN} \end{pmatrix}$ depend on the gauge vector $\mathbf{\Delta} = \{\Delta_1, \dots, \Delta_n\}^T$ as $\Lambda^{EE}(\mathbf{\Delta}) = \Lambda^{EE}(\mathbf{0}) + 2\mathbf{\Delta}^T \mathbf{\Lambda}^{NE}(\mathbf{0}) + \mathbf{\Delta}^T \mathbf{\Lambda}^{NN} \mathbf{\Delta}$ and $\mathbf{\Lambda}^{NE}(\mathbf{\Delta}) = \mathbf{\Lambda}^{NE}(\mathbf{0}) + \mathbf{\Lambda}^{NN}\mathbf{\Delta}$. For the unit mass vector $\widehat{\mathbf{m}} = \frac{(m_1,\dots,m_n)^T}{\sqrt{\sum_{i=1}^n m_i^2}}$ of an $n$-species system, momentum conservation $\mathbf{P}_{tot} = \mathbf{0}$ in equilibrium MD leads to $\mathbf{\Lambda}^{NN}\widehat{\mathbf{m}} = \mathbf{0}$ and $\mathbf{\Lambda}^{EN}\widehat{\mathbf{m}} = 0$. Therefore, $\mathbf{\Lambda}^{NN}$ is singular and positive semidefinite. Introducing a non-singular matrix $\mathbf{\Lambda}^{NN}_\epsilon = \mathbf{\Lambda}^{NN} + \epsilon \widehat{\mathbf{m}} \widehat{\mathbf{m}}^T$ with a nonzero $\epsilon$, we can decompose $\mathbf{\Lambda}$ as $\mathbf{\Lambda} = \begin{pmatrix} 1 & \widetilde{\mathbf{\Lambda}}^{EN}\mathbf{\Lambda}^{NN^{-1}}_\epsilon \\ \mathbf{0} & \mathbf{I} \end{pmatrix} \begin{pmatrix} \widetilde{\Lambda}^{EE} - \widetilde{\mathbf{\Lambda}}^{EN}\mathbf{\Lambda}^{NN^{-1}}_\epsilon\widetilde{\mathbf{\Lambda}}^{NE} & \mathbf{0} \\ \mathbf{0} & \mathbf{\Lambda}^{NN} \end{pmatrix}\begin{pmatrix} 1 & \mathbf{0} \\ \mathbf{\Lambda}^{NN^{-1}}_\epsilon\widetilde{\mathbf{\Lambda}}^{NE} & \mathbf{I} \end{pmatrix}$. All elements involving $\mathbf{\Lambda}^{NN}_\epsilon$ are independent of $\epsilon$ due to the orthogonality relation $\widetilde{\mathbf{\Lambda}}^{EN}\widehat{\mathbf{m}} = \widehat{\mathbf{m}}^T\widetilde{\mathbf{\Lambda}}^{NE} = 0$. From $\widetilde{\Lambda}^{EE} = \Lambda^{EE}(\mathbf{\Delta})$ and $\widetilde{\mathbf{\Lambda}}^{NE} = \mathbf{\Lambda}^{NE}(\mathbf{\Delta})$, it can be shown that $\widetilde{\Lambda}^{EE} - \widetilde{\mathbf{\Lambda}}^{EN}\mathbf{\Lambda}^{NN^{-1}}_\epsilon\widetilde{\mathbf{\Lambda}}^{NE}$ yields the same value, regardless of the choice of $\mathbf{\Delta}$ (ref. [23]). This invariant scalar, which we denote as $\Lambda^{EE}_{min}$, corresponds to the minimum of the quadratic function $\Lambda^{EE}(\mathbf{\Delta})$. Defining the vector

$$\widetilde{\boldsymbol{\varepsilon}} = \mathbf{\Lambda}^{NN^{-1}}_\epsilon \widetilde{\mathbf{\Lambda}}^{NE}, \qquad (9)$$

the decomposition simplifies to $\mathbf{\Lambda} = \begin{pmatrix} 1 & \tilde{\boldsymbol{\varepsilon}}^T \\ 0 & \mathbf{I} \end{pmatrix} \begin{pmatrix} \Lambda_{min}^{EE} & 0 \\ 0 & \mathbf{\Lambda}^{NN} \end{pmatrix} \begin{pmatrix} 1 & 0 \\ \tilde{\boldsymbol{\varepsilon}} & \mathbf{I} \end{pmatrix}$. The component $\tilde{\varepsilon}_i$ in $\tilde{\boldsymbol{\varepsilon}} = (\tilde{\varepsilon}_1, \tilde{\varepsilon}_2, \dots, \tilde{\varepsilon}_n)^T$ can be interpreted as a single-particle energy of species $i$, dynamically averaged within the GK formalism. Applying this matrix identity to equation (8), we obtain

$$\begin{bmatrix} \tilde{J}_x^E - \tilde{\boldsymbol{\varepsilon}}^T \boldsymbol{J}_x^N \\ \boldsymbol{J}_x^N \end{bmatrix} = \begin{pmatrix} \Lambda_{min}^{EE} & 0 \\ \mathbf{\Lambda}^{NN}(\tilde{\boldsymbol{\varepsilon}} - \tilde{\boldsymbol{h}}) & \mathbf{\Lambda}^{NN} \end{pmatrix} \begin{bmatrix} -\frac{1}{k_B T^2} \frac{\partial T}{\partial x} \\ -\frac{1}{k_B T}\left(\frac{\partial \boldsymbol{\mu}}{\partial x}\right)_T \end{bmatrix}, \quad (10)$$

where $\tilde{\boldsymbol{\varepsilon}} = \boldsymbol{\varepsilon}(\boldsymbol{\Delta})$ transforms as $\boldsymbol{\varepsilon}(\boldsymbol{\Delta}) = \boldsymbol{\varepsilon}(0) + \boldsymbol{\Delta} - (\widehat{\boldsymbol{m}}^T \boldsymbol{\Delta})\widehat{\boldsymbol{m}}$, but the last term vanishes in equation (10) since it couples to $\mathbf{\Lambda}^{NN}$ and $\mathbf{\Lambda}^{NN} \widehat{\boldsymbol{m}} = \mathbf{0}$. The renormalized partial enthalpy $\tilde{h}_i = h_i(\boldsymbol{\Delta})$ transforms as $\boldsymbol{h}(\boldsymbol{\Delta}) = \boldsymbol{h}(0) + \boldsymbol{\Delta}$, which follows from equation (12), where $\overline{U}_i \mapsto \overline{U}_i + \Delta_i$ and $\bar{\sigma}_i \mapsto \bar{\sigma}_i$ under the global gauge transformation. As a result, the gauge-dependences in $\tilde{\boldsymbol{\varepsilon}}$ and $\tilde{\boldsymbol{h}}$ within $\mathbf{\Lambda}^{NN}(\tilde{\boldsymbol{\varepsilon}} - \tilde{\boldsymbol{h}})$ cancel out, making the number flux $\boldsymbol{J}_x^N$ invariant.

Given that $\Lambda_{min}^{EE}$ is an invariant, $\tilde{J}_x^E - \tilde{\boldsymbol{\varepsilon}}^T \boldsymbol{J}_x^N = -\frac{\Lambda_{min}^{EE}}{k_B T^2} \frac{\partial T}{\partial x}$ in equation (10) is a gauge-invariant form of energy flux. However, it cannot be identified as heat flux $J_x^H$ since the matrix in equation (10) is asymmetric. According to Onsager's reciprocal relation[28,29], the matrix relating $J_x^H$ and $\boldsymbol{J}_x^N$ to external perturbations should have identical cross-dissipative coefficients. Therefore, $\begin{bmatrix} J_x^H \\ \boldsymbol{J}_x^N \end{bmatrix} = \begin{pmatrix} \Lambda_{min}^{EE} + C & (\tilde{\boldsymbol{\varepsilon}} - \tilde{\boldsymbol{h}})^T \mathbf{\Lambda}^{NN} \\ \mathbf{\Lambda}^{NN}(\tilde{\boldsymbol{\varepsilon}} - \tilde{\boldsymbol{h}}) & \mathbf{\Lambda}^{NN} \end{pmatrix} \begin{bmatrix} -\frac{1}{k_B T^2} \frac{\partial T}{\partial x} \\ -\frac{1}{k_B T}\left(\frac{\partial \boldsymbol{\mu}}{\partial x}\right)_T \end{bmatrix}$, which contains a gauge-invariant scalar $C$ as a degree of freedom. By selecting $C = (\tilde{\boldsymbol{\varepsilon}} - \tilde{\boldsymbol{h}})^T \mathbf{\Lambda}^{NN}(\tilde{\boldsymbol{\varepsilon}} - \tilde{\boldsymbol{h}})$, we finally get $J_x^H = (\tilde{J}_x^E - \tilde{\boldsymbol{\varepsilon}}^T \boldsymbol{J}_x^N) + (\tilde{\boldsymbol{\varepsilon}} - \tilde{\boldsymbol{h}})^T \boldsymbol{J}_x^N = \tilde{J}_x^E - \tilde{\boldsymbol{h}}^T \boldsymbol{J}_x^N$, which now depends on the external perturbations only implicitly. With this symmetry-dictated definition of $J_x^H$, we obtain

$$\begin{bmatrix} J_x^H \\ \boldsymbol{J}_x^N \end{bmatrix} = \begin{pmatrix} \Lambda_{min}^{EE} + (\tilde{\boldsymbol{\varepsilon}} - \tilde{\boldsymbol{h}})^T \mathbf{\Lambda}^{NN}(\tilde{\boldsymbol{\varepsilon}} - \tilde{\boldsymbol{h}}) & (\tilde{\boldsymbol{\varepsilon}} - \tilde{\boldsymbol{h}})^T \mathbf{\Lambda}^{NN} \\ \mathbf{\Lambda}^{NN}(\tilde{\boldsymbol{\varepsilon}} - \tilde{\boldsymbol{h}}) & \mathbf{\Lambda}^{NN} \end{pmatrix} \begin{bmatrix} -\frac{1}{k_B T^2} \frac{\partial T}{\partial x} \\ -\frac{1}{k_B T}\left(\frac{\partial \boldsymbol{\mu}}{\partial x}\right)_T \end{bmatrix}. \quad (11)$$

**Local gauge invariance.** As emphasized above, for the internal consistency of the theory, the partial enthalpy $h_i$ should be properly renormalized. Given that $E = \sum_{j=1}^{n} N_j \left(\overline{U}_j + \frac{3}{2} k_B T\right)$ and $P = \frac{1}{\Omega}\sum_{j=1}^{n} N_j(\bar{\sigma}_j + k_B T)$[39], we obtain the enthalpy $H = E + P\Omega = \sum_{j=1}^{n} N_j\left(\bar{U}_j + \bar{\sigma}_j + \frac{5}{2}k_B T\right)$. Comparing this with the sum rule $H = \sum_{j=1}^{n} N_j h_j$ suggests a renormalized form of $\tilde{h}_i$:

$$\tilde{h}_i = \bar{U}_i + \bar{\sigma}_i + C_i, \qquad (12)$$

where $C_i$ is a gauge-invariant (i.e., MLP-independent) constant for species $i$. Interestingly, the single-particle energy $\tilde{\varepsilon}_i$ in equation (9) also takes the form $\tilde{\varepsilon}_i = \bar{U}_i + \bar{\sigma}_i + C_i'$ for another constant $C_i'$ (Methods). Figure 2c demonstrates the linear relation between $\tilde{\varepsilon}_{Cu}$ and $\bar{U}_{Cu} + \bar{\sigma}_{Cu}$ for $\alpha$-Cu$_2$S. Therefore, if equation (12) holds, the identical gauge dependence of $\tilde{\boldsymbol{\varepsilon}}$ and $\tilde{\boldsymbol{h}}$ ensures the local gauge invariance of the heat flux $J_x^H$ in equation (11).

The essence of renormalizing partial enthalpies lies in ensuring that both $\tilde{h}_i$ and $\tilde{\varepsilon}_i$ derive their arbitrariness *solely* from the MLP function $U_I = U_{\text{MLP}}^{\{N_i\}}(x_I)$ (equation (5)), which is trained for a stoichiometric system with composition $\{N_i\}$. To derive the expression for $\tilde{h}_i$, we first rewrite the partial enthalpy $h_i = \left(\frac{\partial H}{\partial N_i}\right)_{T,P}$ as $h_i = \frac{H(N_i+1) - H(N_i-1)}{2}$, where $H(N_i \pm 1)$ represents the enthalpy at given $T$ and $P$ for a system in which $N_i$ has been increased or decreased by one. The renormalized $\tilde{h}_i$ is then obtained by replacing $H(N_i \pm 1)$ with $\tilde{H}(N_i \pm 1) = \sum_{j=1}^{n} N_j\left(\bar{U}_j^{\{N_i\}} + \bar{\sigma}_j^{\{N_i\}} + \frac{5}{2}k_B T\right)$, where $\bar{U}_j^{\{N_i\}}$ and $\bar{\sigma}_j^{\{N_i\}}$ (equation (6)) are computed for systems with $N_i \pm 1$, using the atomic energies obtained from the function $U_I = U_{\text{MLP}}^{\{N_i\}}(x_I)$ as the *single* source of arbitrariness. Through this renormalization, $\tilde{h}_i$ satisfies equation (12), as shown in Methods and numerically confirmed in Fig. 2c for $\alpha$-Cu$_2$S. Consequently, we obtain consistent TE coefficients across different MLPs (Fig. 2d).

**Thermoelectricity.** For ionic TE, equation (11) can be simplified by introducing a symmetry on charge transport. In equation (2), the charge flux $\boldsymbol{J}^C(t)$ is defined in terms of the real-valued, time-dependent tensor $\boldsymbol{Z}_I^*(t)$ from an equilibrium MD trajectory $\boldsymbol{R}_I(t)$. Instead, motivated by Thouless' theorem[42–44] on adiabatic charge transport, we may use $\boldsymbol{J}^{C'}(t) = \frac{1}{\Omega}\sum_I q_{s(I)}\boldsymbol{v}_I(t)$, where $q_i$ is an integer-valued, time-independent scalar charge for species $i$. Recently, Grasselli and Baroni[16] showed that if strong adiabaticity holds, $\boldsymbol{J}^C(t)$ and $\boldsymbol{J}^{C'}(t)$ yield the same electrical conductivity $\sigma = \frac{\Lambda^{CC}}{k_B T}$. This remarkable symmetry allows us to write the charge flux as $J_x^C = \boldsymbol{q}^T J_x^N$ with the topologically quantized charges $\boldsymbol{q} = (q_1, q_2, ..., q_n)^T$. Then, using the relation $\left(\frac{\partial \mu}{\partial x}\right)_T = \boldsymbol{q}\frac{\partial V}{\partial x}$ for a voltage gradient $\frac{\partial V}{\partial x}$, equation (11) transforms into

$$\begin{bmatrix} J_x^H \\ J_x^C \end{bmatrix} = \begin{pmatrix} \Lambda_{min}^{EE} + (\tilde{\boldsymbol{\varepsilon}} - \tilde{\boldsymbol{h}})^T \Lambda^{NN}(\tilde{\boldsymbol{\varepsilon}} - \tilde{\boldsymbol{h}}) & (\tilde{\boldsymbol{\varepsilon}} - \tilde{\boldsymbol{h}})^T \Lambda^{NN} \boldsymbol{q} \\ \boldsymbol{q}^T \Lambda^{NN}(\tilde{\boldsymbol{\varepsilon}} - \tilde{\boldsymbol{h}}) & \boldsymbol{q}^T \Lambda^{NN} \boldsymbol{q} \end{pmatrix} \begin{bmatrix} -\frac{1}{k_B T^2}\frac{\partial T}{\partial x} \\ -\frac{1}{k_B T}\frac{\partial V}{\partial x} \end{bmatrix}. \quad (13)$$

To provide a geometrical interpretation of equation (13), we introduce an $(n-1)$-dimensional vector space for an $n$-species system (Fig. 1b). In this space, a vector $|v\rangle$ is defined as $|v\rangle = \sqrt{\frac{\Lambda^{NN}}{\lambda}}\boldsymbol{v}$, where $\frac{\Lambda^{NN}}{\lambda}$ is a positive semi-definite matrix normalized by the largest eigenvalue $\lambda$ of $\Lambda^{NN}$. The inner product $\langle v|w\rangle$ between two vectors $|v\rangle$ and $|w\rangle$ is given by $\langle v|w\rangle = \boldsymbol{v}^T \frac{\Lambda^{NN}}{\lambda}\boldsymbol{w}$. Since $\langle \hat{\boldsymbol{m}}|v\rangle = 0$ due to $\Lambda^{NN}\hat{\boldsymbol{m}} = \boldsymbol{0}$ for the unit mass vector $\hat{\boldsymbol{m}}$, the vector space has only $n-1$ orthonormal basis vectors. In the open-circuit condition $J_x^C = 0$ in equation (13), we obtain the ionic Seebeck coefficient S from the definition $\frac{\partial V}{\partial x} = -S\frac{\partial T}{\partial x}$ and the ionic thermal conductivity κ from $J_x^H = -\kappa\frac{\partial T}{\partial x}$. With the length $\|v\| = \sqrt{\langle v|v\rangle}$ and the angle $\theta = \cos^{-1}\frac{\langle v|w\rangle}{\|v\|\|w\|}$, the ionic S has a simple geometric representation (Fig. 1b):

$$S = \frac{\cos\theta_{TE}}{T}\frac{\|\boldsymbol{\varepsilon}-\boldsymbol{h}\|}{\|\boldsymbol{q}\|}, \quad (14)$$

where $\theta_{TE}$ denotes the angle between $|\tilde{\boldsymbol{\varepsilon}} - \tilde{\boldsymbol{h}}\rangle$ and $|\boldsymbol{q}\rangle$. The ionic κ also depends on $\theta_{TE}$ as

$$\kappa = \frac{1}{k_B T^2}(\Lambda_{min}^{EE} + \lambda\|\boldsymbol{\varepsilon}-\boldsymbol{h}\|^2 \sin^2\theta_{TE}), \quad (15)$$

indicating that the previous result[23,27], $\kappa = \frac{\Lambda_{min}^{EE}}{k_B T^2}$, serves as a lower bound of $\kappa \geq \frac{\Lambda_{min}^{EE}}{k_B T^2}$ (see below). Under no temperature gradient, we get the electrical conductivity $\sigma = \frac{\lambda\|\boldsymbol{q}\|^2}{k_B T}$ and the Peltier coefficient $\pi = TS$ from $J_x^H = \pi J_x^C$.

For two-species systems such as $Cu_2S$, $Li_3N$, and molten KCl, the vector space in Fig. 1b is one-dimensional, restricting $\theta_{TE}$ to 0 or $\pi$. Another example is molten $KNO_3$, which in a coarse-grained model, can be regarded as a two-species system consisting of $K^+$ cations and $NO_3^-$ anions, and thus $\theta_{TE} = 0$ or $\pi$. In this case, $S = \frac{1}{T}\frac{\pm\|\boldsymbol{\varepsilon}-\boldsymbol{h}\|}{\|\boldsymbol{q}\|}$ from equation (14), analogous to the electronic $S_e$ in equation (1). The ionic Seebeck effect is p-type (S > 0) when $\theta_{TE} = 0$ and n-type when $\theta_{TE} = \pi$. In contrast, multi-species systems such as KNO$_3$−NaNO$_3$ mixtures do not have this angle restriction. The power factor $\sigma S^2$ depends on the angle $\theta_{TE}$ as $\sigma S^2 = \frac{\lambda \|\boldsymbol{\varepsilon}-\boldsymbol{h}\|^2 \cos^2 \theta_{TE}}{k_B T^3}$, which decreases when $\theta_{TE}$ deviates from 0 or $\pi$. Using the power factor, the κ in equation (15) can be rewritten as $\kappa = \kappa_0 - \sigma S^2 T$, where $\kappa_0 = \frac{\Lambda_{min}^{EE} + \lambda \|\boldsymbol{\varepsilon}-\boldsymbol{h}\|^2}{k_B T^2}$ denotes the intrinsic thermal conductivity, and the correction term $-\sigma S^2 T$ reflects the reduction in heat flow due to the TE coupling. When $\theta_{TE} = 0$ or $\pi$, the second term in $\kappa_0$ and the TE correction term cancel out, leaving the simplified expression $\kappa = \frac{\Lambda_{min}^{EE}}{k_B T^2}$.

**Numerical experiments**

We present numerical results for two representative systems $\alpha$-Cu$_2$S and $\alpha$-Li$_3$N, obtained using MD trajectories in MLP-driven MD simulations (Methods). Our first example is $\alpha$-Cu$_2$S, the high-temperature cubic phase characterized by the liquid-like motions of Cu cations within a hexagonal lattice of S anions[27,45]. From 26 million MD steps for a 4116-atom supercell at $T$ = 800 K, we obtain the ionic Seebeck coefficient S = $0.446 \pm 0.197\ k_B/e$ ($1 k_B/e = 86\ \mu V/K$). Compared to other transport coefficients (e.g., κ and σ), the ionic S converges slowly with increasing MD steps. In experiments[31], the sample is unintentionally hole-doped, meaning that both Cu ions and holes contribute to the total Seebeck coefficient $S_{tot}$. Compared to the measured value $S_{tot} \approx 4\ k_B/e$ in ref. [31], the ionic contribution S is relatively small, in consistent with the consensus that the TE in Cu$_2$S is mainly electronic in origin[30,31].

The solid-state electrolyte, $\alpha$-Li$_3$N, consists of alternating layers of Li$_2$N and pure Li (Fig. 3a). For this anisotropic ion conductor, we performed a MD simulation using a 4000-atom supercell at $T$ = 673 K for 35 million MD steps. Our TE theory predicts a negative Seebeck coefficient $S_{xy} = -4.14 \pm 0.58\ k_B/e$ for conduction along the $x$ (or $y$) axis and a positive coefficient $S_z = 3.13 \pm 1.09\ k_B/e$ for conduction along the $z$ axis. Thus, the single system, Li$_3$N, can exhibit both n-type and p-type ionic TE depending on the transport direction. Given that Li$_3$N is an ionic conductor for Li cations, the negative $S_{xy}$ was unexpected, suggesting that the in-plane Li-ion transport involves Li vacancies. Indeed, previous studies[46,47] showed that (i) Li vacancies are created in the Li$_2$N layers concomitantly with Li split interstitials and (ii) these vacancies facilitate the relatively fast diffusion of Li ions in the Li$_2$N layers (top panel in Fig. 3b). In contrast, the interlayer motion of Li ions is mediated by the formation of Li split interstitials at the Li sites linking the Li$_2$N layers[46] (bottom panel), resulting in the positive $S_z$ for out-of-plane transport.

**Conclusions**

We have developed an energy-gauge invariant theory of ionic TE that is universally applicable, independent of specific ion transport mechanisms. The ambiguity arising from ill-defined atomic energies is resolved by renormalizing partial enthalpies in terms of the atomic energies themselves. When combined with ab initio-quality MLP-driven MD simulations, our theory provides a reliable computational framework for determining ionic Seebeck and other transport coefficients in real materials. Notably, our numerical simulations reveal that the solid-state electrolyte $\alpha$-Li$_3$N exhibits distinct Seebeck coefficient polarities depending on the direction of Li-ion transport: (i) $S_{xy} < 0$ for in-plane transport and (ii) $S_z > 0$ for out-of-plane transport. These findings suggest that Seebeck measurements could serve as a powerful tool for probing the dynamically organized motion of ions in ionic conductors, analogous to Hall measurements providing useful information (e.g., carrier polarity) on electronic properties.

**Acknowledgements**

This work was supported by the National Research Foundation grant (No. NRF-2022M3D1A1026816) funded by the Ministry of Science and ICT of Korea. This research used the resources of the DGIST Supercomputing and Bigdata Center.

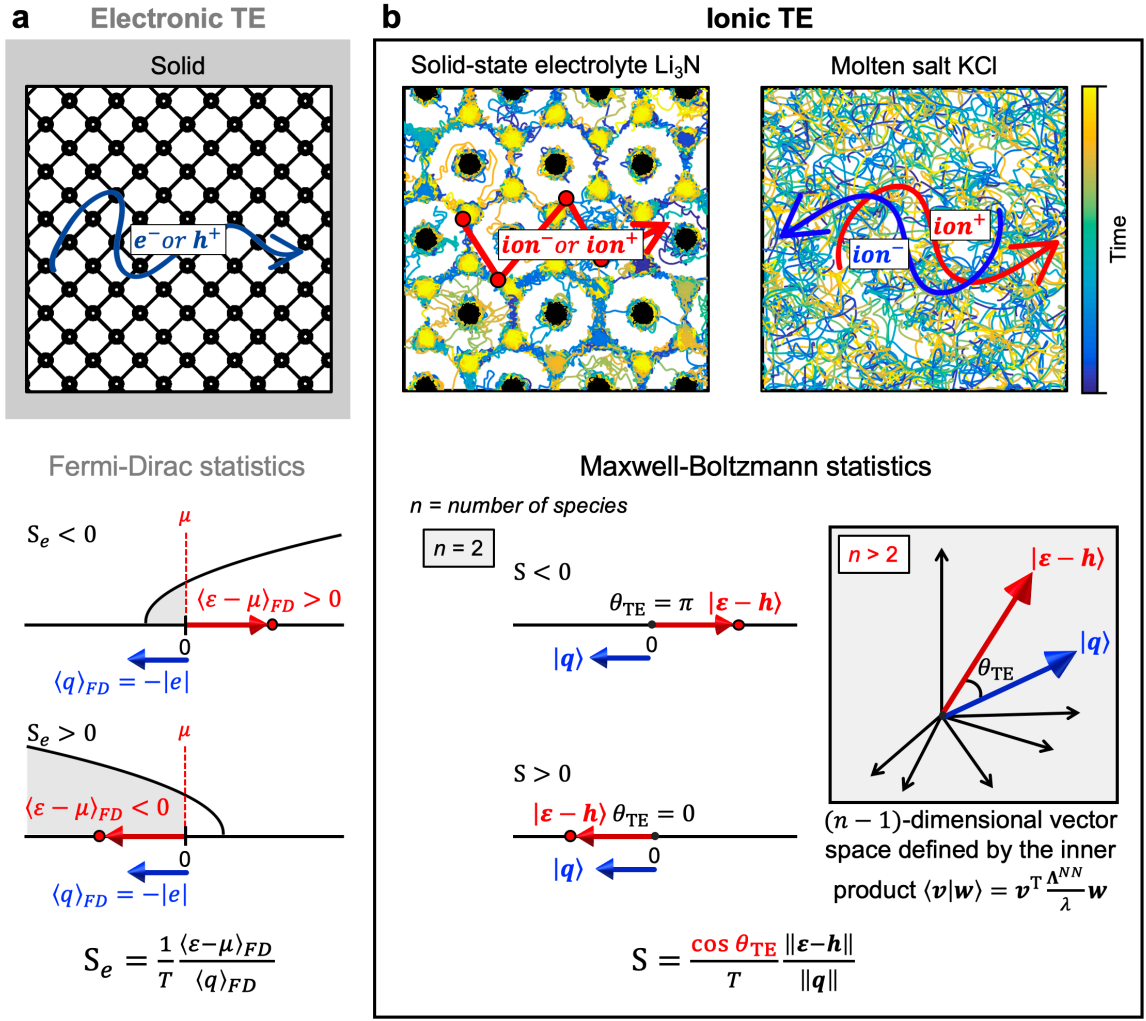

**Fig. 1. Electronic vs ionic thermoelectricity.** (a) Electronic thermoelectricity in solid metals or doped semiconductors, described by the Mott relations and the Fermi-Dirac statistics. (b) Ionic thermoelectricity in electronically gapped ionic liquids, such as solid-state electrolytes and molten salts, governed by the Maxwell-Boltzmann statistics. The ionic Seebeck coefficient S in equation (14) has a simple geometric representation (lower right panel). When $\theta_{TE} = 0$ or $\pi$, as in two-species systems with $n = 2$, the ionic S has the same representation as the electronic $S_e$. However, this analogy does not hold when the angle $\theta_{TE}$ deviates from 0 or $\pi$.

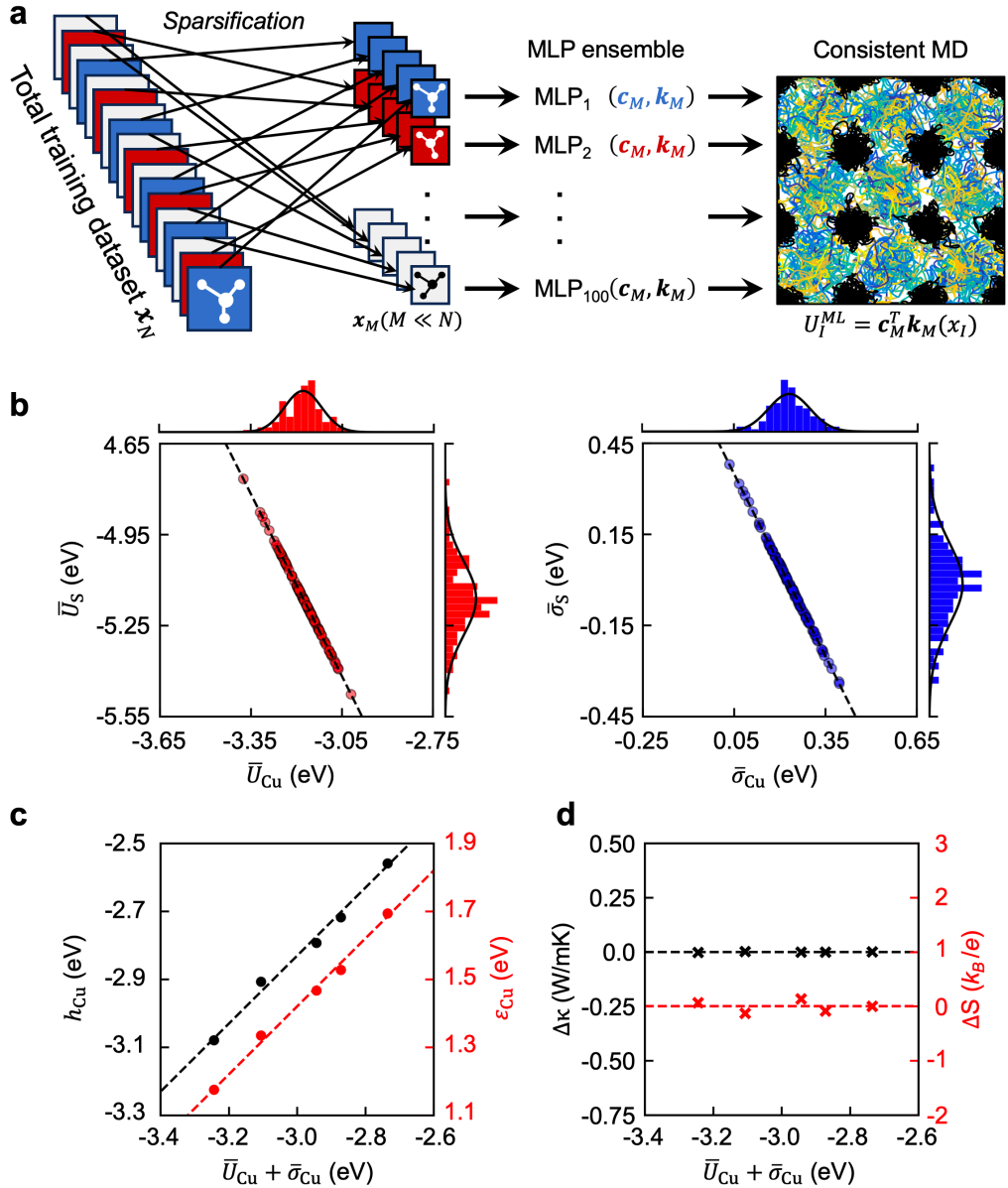

**Fig. 2. Gauge invariance of ionic thermoelectricity.** (a) Generation of the ensemble of 100 MLPs through the sparse GPR fitting of local atomic energies $U_I$ and (b) Gaussian distributions of the time averages, $\overline{U}_i$ and $\overline{\sigma}_i$, for $i$ = Cu and S in the system $\alpha$-Cu$_2$S, obtained with the MLP ensemble for a given MD trajectory at $T$ = 800 K. (c) The transformation rules for the renormalized partial enthalpy $\tilde{h}_{Cu}$ (left) and the single-particle energy $\tilde{\varepsilon}_{Cu}$ (right), demonstrated with five of the MLPs in the ensemble having different values of $\overline{U}_{Cu} + \overline{\sigma}_{Cu}$. (d) Local gauge invariance of the thermal conductivity κ and the Seebeck coefficient S, numerically demonstrated using the five MLPs selected in (c). MLP-dependent variations, Δκ (left) and ΔS (right), are plotted.

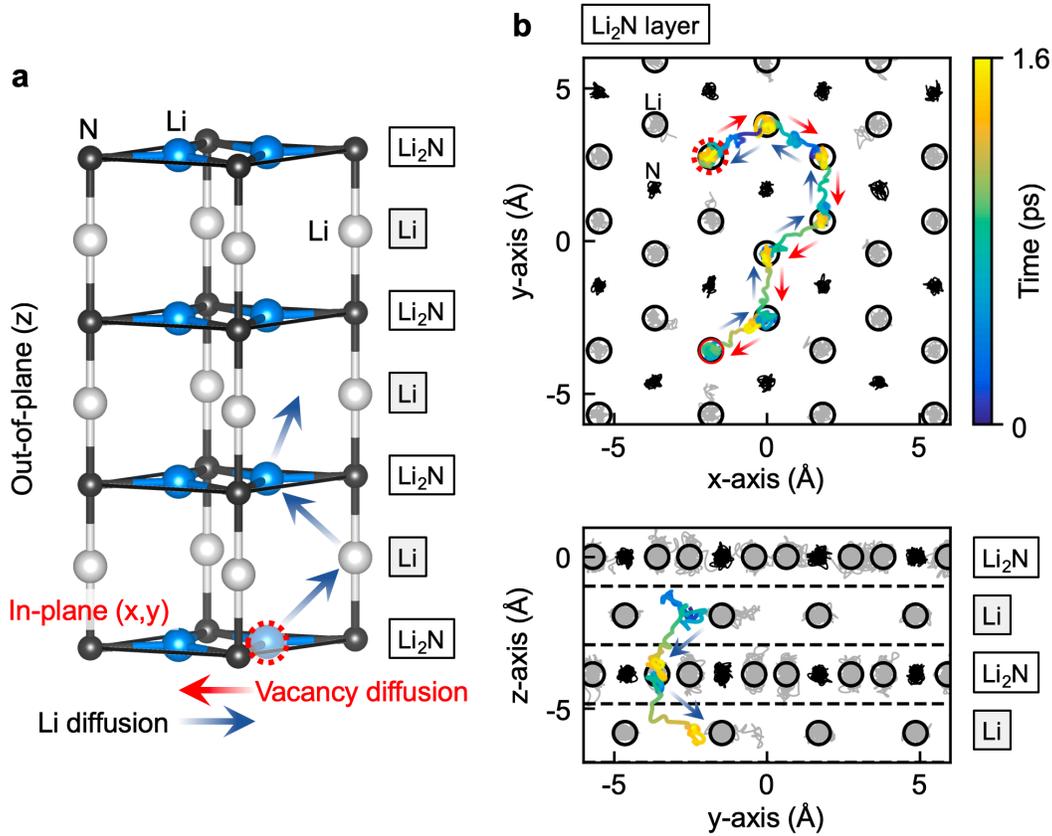

**Fig. 3. Highly anisotropic ionic thermoelectricity in Li₃N.** (a) Atomic structure of the solid-sate electrolyte $\alpha$-Li$_3$N, consisting of alternating layers of Li$_2$N and pure Li. (b) MD trajectories at $T =$ 673 K illustrating Li vacancy diffusion within the Li$_2$N layer (top) and interlayer Li-ion diffusion mediated by the formation of Li split interstitials at the Li sites linking the Li$_2$N layers (bottom). Different transport mechanisms for the in-plane and out-of-plane motions lead to distinct polarities of the ionic Seebeck coefficient depending on the transport direction of Li ions in the solid-sate electrolyte $\alpha$-Li$_3$N: (i) $S_{xy} < 0$ for conduction along the $x$ (or $y$) axis and (ii) $S_z > 0$ for conduction along the $z$ axis.

## Methods

**MLP ensemble generation.** A machine-learned potential (MLP) serves a dual purpose in our study. First, it acts as an *ab initio*-quality force field, enabling large-scale MD simulations. In this approach, the system's potential energy $U$ is decomposed into "atomic" energies, $U_I = U^{\text{MLP}}(x_I)$, based on the local atomic environments $x_I$ around individual atoms $I$. The function $U^{\text{MLP}}(x_I)$ is approximated via supervised ML to satisfy the sum rule $U = \sum_{I=1}^{N} U_{\text{MLP}}(x_I)$ for systems included in the training data. Consequently, the total energy of an arbitrarily large system can be obtained by summing the atomic energies $U^{\text{MLP}}(x_I)$. Second, these atomic energies are used to compute the energy flux $\boldsymbol{J}^E(t)$ in equilibrium MD simulations (equations (3) and (4)).

To generate an ensemble of 100 MLPs for $\alpha$-Cu$_2$S, we employed a mixed kernel model in Gaussian Approximation Potential, which combines a squared-exponential kernel for two-body descriptors and the smooth overlap of atomic positions (SOAP) kernel. The MLPs in the ensemble were independently trained via sparse Gaussian process regression fitting of $U^{\text{MLP}}(x_I)$, using the same kernel and training data prepared by density function theory calculations. As a result, all MLPs satisfy the sum rule $U = \sum_{I=1}^{N} U_{\text{MLP}}(x_I)$ with similar accuracy. However, the functions $U^{\text{MLP}}(x_I)$, based on the mixed kernel model, vary substantially across the MLP ensemble (Fig. 2).

To obtain the results in Figs. 2b–2d, we performed a constant-temperature MD simulation at $T$ = 800 K using a 3888-atom supercell of the high-temperature cubic phase $\alpha$-Cu$_2$S. From the ensemble, a single MLP was selected as the ML force field[48] to generate a single MD trajectory consisting of 3 million MD steps. Using the same MD trajectory, we computed the local atomic energies $U_I = U^{\text{MLP}}(x_I)$ for the 100 MLPs in the ensemble. Since the time-averaged quantities $\overline{U}_i$ and $\overline{\sigma}_i$ (equation (6)) converge quickly with simulation time $t$, only 1000 MD steps were required to generate the distributions of $\overline{U}_i$ and $\overline{\sigma}_i$ in Fig. 2b. In contrast, the full MD trajectory was used to obtain the results in Figs. 2c and 2d.

**Renormalization of partial enthalpies.** The gauge-invariant constant $C_i$ in equation (12) can be obtained using either the $N\Omega T$ ensemble (as described below) or the $NPT$ ensemble for equilibrium MD simulations. Our analysis, based on the central limit theorem, shows that the $N\Omega T$ ensemble provides statistically more converged results for a given MD simulation length compared to the $NPT$ ensemble. To express the partial enthalpy $h_i$ in terms of quantities that can be obtained using the $N\Omega T$ ensemble, we rewrite $h_i = \left(\frac{\partial H}{\partial N_i}\right)_{T,P} = \left(\frac{\partial E}{\partial N_i}\right)_{T,P} + P\left(\frac{\partial \Omega}{\partial N_i}\right)_{T,P}$ for a stoichiometric system with composition $\{N_1, \ldots, N_n\}$ as

$$h_i = \left(\frac{\partial E}{\partial N_i}\right)_{T,\Omega} - T\frac{\left(\frac{\partial P}{\partial T}\right)_\Omega \left(\frac{\partial P}{\partial N_i}\right)_{T,\Omega}}{\left(\frac{\partial P}{\partial \Omega}\right)_T} \qquad (16)$$

using the relations $\left(\frac{\partial E}{\partial N_i}\right)_{T,P} = \left(\frac{\partial E}{\partial N_i}\right)_{T,\Omega} + \left(\frac{\partial E}{\partial \Omega}\right)_T \left(\frac{\partial \Omega}{\partial N_i}\right)_{T,P}$, $P = -\left(\frac{\partial E}{\partial \Omega}\right)_T + T\left(\frac{\partial S}{\partial \Omega}\right)_T$, and Maxwell's relation $\left(\frac{\partial S}{\partial \Omega}\right)_T = \left(\frac{\partial P}{\partial T}\right)_\Omega$. To obtain the renormalized $\tilde{h}_i$, we replace the partial derivatives in equation (16) with $\left(\frac{\partial E}{\partial N_i}\right)_{T,\Omega} \to \left(\frac{\partial \tilde{E}}{\partial N_i}\right)_{T,\Omega} = \frac{\tilde{E}(N_i+1)-\tilde{E}(N_i-1)}{2}$ and $\left(\frac{\partial P}{\partial N_i}\right)_{T,\Omega} \to \left(\frac{\partial \tilde{P}}{\partial N_i}\right)_{T,\Omega} = \frac{\tilde{P}(N_i+1)-\tilde{P}(N_i-1)}{2}$, where $\tilde{E}$ and $\tilde{P}$ for $N_i \pm 1$ are defined as $\tilde{E} = \sum_{j=1}^{n} N_j \left(\bar{U}_j^{\{N_i\}} + \frac{3}{2}k_B T\right)$ and $\tilde{P} = \frac{1}{\Omega}\sum_{j=1}^{n} N_j \left(\bar{\sigma}_j^{\{N_i\}} + k_B T\right)$ for a system in which $N_i$ has been increased or decreased by one. In these equations, $\bar{U}_j^{\{N_i\}}$ and $\bar{\sigma}_j^{\{N_i\}}$ (equation (6)) are computed using the function $U_I = U_{MLP}^{\{N_i\}}(x_I)$ (equation (5)), which is trained for the original system with $N_i$.

Given the expressions for $\tilde{E}$ and $\tilde{P}$, one might expect the finite-difference expressions for $\left(\frac{\partial \tilde{E}}{\partial N_i}\right)_{T,\Omega}$ and $\left(\frac{\partial \tilde{P}}{\partial N_i}\right)_{T,\Omega}$ to include gauge-dependences as $\bar{U}_i^{\{N_i\}}$ and $\frac{1}{\Omega}\bar{\sigma}_i^{\{N_i\}}$, respectively. However, this is not the case. If a particle of species $i$ traverses an 'imaginary' boundary (e.g., a $yz$ plane) between two cubic supercells, each containing $N_i$, the energy change $\tilde{E}(N_i+1) - E(N_i)$ on one side arises in part from the energy flux $\mathcal{J}_x^E = \frac{1}{\Omega}\sum_{I,\beta} E_{Ix\beta}^* v_{I\beta}$ at the boundary. On average, this flux contributes the time average of $E_{Ixx}^*(t)$ for species $i$:

$$\bar{E}_i^* = \frac{1}{N_i}\sum_{I \in i}\langle E_{Ixx}^*(t)\rangle = \bar{U}_i + \bar{\sigma}_i + \frac{3}{2}k_B T. \qquad (17)$$

Thus, the gauge-dependence in $\left(\frac{\partial \tilde{E}}{\partial N_i}\right)_{T,\Omega}$ enters as $\bar{U}_i^{\{N_i\}} + \bar{\sigma}_i^{\{N_i\}}$, rather than solely as $\bar{U}_i^{\{N_i\}}$. Indeed, numerical calculations confirm this gauge dependence for $\left(\frac{\partial \tilde{E}}{\partial N_i}\right)_{T,\Omega}$, while $\left(\frac{\partial \tilde{P}}{\partial N_i}\right)_{T,\Omega}$ remains gauge-invariant. Consequently, the renormalized $\tilde{h}_i$ possesses the gauge dependence as $\tilde{h}_i = \bar{U}_i^{\{N_i\}} + \bar{\sigma}_i^{\{N_i\}} + C_i$ (equation (12)).

**Gauge transformation of single-particle energies.** As discussed in the main text, the atomic energies $U_I$ of two MLPs, P$_1$ and P$_2$, in the ensemble are not simply related by the global gauge transformation in equation (7). Nonetheless, we can always find the energy gauges $\bar{\Delta}_i$ that satisfy the following equations for $\boldsymbol{\Lambda} = \begin{pmatrix} \widetilde{\Lambda}^{EE} & \widetilde{\Lambda}^{EN} \\ \widetilde{\Lambda}^{NE} & \Lambda^{NN} \end{pmatrix}$:

$$\widetilde{\Lambda}^{EE}[\{U_I^{P_1}(t)\}] = \widetilde{\Lambda}^{EE}[\{U_I^{P_2}(t) + \bar{\Delta}_{s(I)}\}], \quad (18a)$$
$$\widetilde{\Lambda}^{NE}[\{U_I^{P_1}(t)\}] = \widetilde{\Lambda}^{NE}[\{U_I^{P_2}(t) + \bar{\Delta}_{s(I)}\}] \quad (18b)$$

for two sets of atomic energies calculated using the potentials P$_1$ and P$_2$. For an *n*-species system, we have $n-1$ independent equations for the components $\widetilde{\Lambda}_i^{NE}$ in equation (18b) due to the orthogonality relation $\widehat{\boldsymbol{m}}^{\mathrm{T}}\widetilde{\Lambda}^{NE} = 0$ for the unit mass vector $\widehat{\boldsymbol{m}}$. Therefore, there are in total *n* independent equations for *n* unknowns $\bar{\Delta}_i$.

It turns out that the energy gauges $\bar{\Delta}_i$ in equation (18) can be expressed as

$$\bar{\Delta}_i = \bar{E}_i^{*P_1} - \bar{E}_i^{*P_2} \quad (19)$$

in terms of the gauge-dependent $\bar{E}_i^*$ in equation (17). Consequently, equation (18) implies

$$\boldsymbol{\Lambda}\left[\{U_I^{P_1}(t) - \bar{E}_{s(I)}^{*P_1}\}\right] = \boldsymbol{\Lambda}\left[\{U_I^{P_2}(t) - \bar{E}_{s(I)}^{*P_2}\}\right]. \quad (20)$$

Upon the global gauge transformation $U_I(t) \mapsto U_I(t) - \bar{E}_{s(I)}$, the atomic energy tensor $E_{I\alpha\beta}^*$ in equation (4) transforms as $E_{I\alpha\beta}^*(t) \mapsto E_{I\alpha\beta}^*(t) - \bar{E}_{s(I)}\delta_{\alpha\beta}$, whose time average becomes zero by construction. Hence, the corresponding energy flux $\mathcal{J}_x^E(t)$ captures only the fluctuations in time, leading to the MLP-independent matrix $\boldsymbol{\Lambda}[\{U_I(t) - \bar{E}_{s(I)}\}]$.

A corollary of equation (18) is that the global gauge invariance of $\Lambda_{min}^{EE} = \widetilde{\Lambda}^{EE} - \widetilde{\Lambda}^{EN}\Lambda_\epsilon^{NN^{-1}}\widetilde{\Lambda}^{NE}$ ensures its invariance under the local gauge transformation. From the transformation rule of the single-particle energies $\tilde{\boldsymbol{\varepsilon}}$ under the global gauge transformation, we also find that

$$\tilde{\varepsilon}^{P_1} = \tilde{\varepsilon}^{P_2} + \overline{\Delta} \qquad (21)$$

apart from the term proportional to the mass vector $\widehat{\boldsymbol{m}}$.

Finally, it follows from equations (19) and (21) that $\tilde{\varepsilon}_i^{P_1} - \overline{E}_i^{*P_1} = \tilde{\varepsilon}_i^{P_2} - \overline{E}_i^{*P_2}$, indicating that $\tilde{\varepsilon}_i - \overline{E}_i$ is gauge-invariant (i.e., MLP-independent). Therefore, using equation (17), we obtain

$$\tilde{\varepsilon}_i = \overline{U}_i + \overline{\sigma}_i + C'_i \qquad (22)$$

for an MLP-independent constant $C'_i$.

**Ab initio-quality MD simulations.** For long-time MD simulations of both $\alpha$-Cu$_2$S and $\alpha$-Li$_3$N, we employed kernel-based potentials generated by an on-the-fly ML algorithm, as implemented in VASP[49,50]. For $\alpha$-Cu$_2$S, we used a 4116-atom supercell with periodic boundary conditions. To calculate the $\boldsymbol{\Lambda}$ matrix within the GK formalism, a constant-temperature MD simulation was conducted at $T = 800$ K for 26 million MD steps with $\Delta t = 1$ fs. Similarly, for $\alpha$-Li$_3$N, we performed a MD simulation using a 4000-atom supercell at $T = 673$ K for 35 million MD steps. For each material, additional MD simulations were performed under slightly varied conditions within the $N\Omega T$ ensemble (equation (16)) to calculate the renormalized partial enthalpies $\tilde{h}_i$ using a finite-difference method.